\documentclass[aps,pre,reprint,amsmath,amssymb,superscriptaddress,showpacs]{revtex4-1}

\usepackage{mathtools}
\usepackage{amsmath}    
\usepackage{graphicx}   
\usepackage{verbatim}   
\usepackage{color}      
\usepackage{lipsum}
\usepackage{subfigure}  
\usepackage{hyperref}   

\begin{document}


\title{On the behavior of the Generalized Alignment Index (GALI) method for regular motion in multidimensional Hamiltonian systems}

\author{Henok Moges}
\email{mgshen002@myuct.ac.za}
\affiliation{Department of Mathematics and Applied Mathematics, University of Cape Town, Rondebosch, 7701, Cape Town, South Africa}
\author{Thanos Manos}%
\email{thanos.manos@u-cergy.fr}
\affiliation{Department of Physics, Faculty of Science and Technology, University of Cergy-Pontoise, Cergy-Pontoise, France}
\author{Charalampos Skokos}
\email{haris.skokos@uct.ac.za}
\affiliation{Department of Mathematics and Applied Mathematics, University of Cape Town, Rondebosch, 7701, Cape Town, South Africa}


\begin{abstract}
We investigate the behavior of the Generalized Alignment Index of order $k$ (GALI$_k$) for regular orbits of multidimensional Hamiltonian systems. The GALI$_k$ is an efficient chaos indicator, which asymptotically attains positive values for regular motion when $2\leq k \leq N$, with $N$ being the dimension of the torus on which the motion occurs. By considering several regular orbits in the neighborhood of two typical simple, stable periodic orbits of the  Fermi-Pasta-Ulam-Tsingou (FPUT) $\beta$ model for various values of the system's degrees of freedom, we show that the asymptotic  GALI$_k$ values decrease when the index's order $k$ increases and when the orbit's energy approaches the periodic orbit's destabilization energy where the stability island vanishes, while they increase when the considered regular orbit moves further away from the periodic one for a fixed energy. In addition, performing extensive numerical simulations we show that the index's behavior does not depend on the choice of the initial deviation vectors needed for its evaluation.
\end{abstract}

\maketitle


\section{Introduction}\label{sec:intro}

Dynamical systems constitute a rather substantial research area with many different applications in several fields of physics. In particular, Hamiltonian systems  of many degrees of freedom (dof) have been broadly used for the study and better understanding of energy transport and equipartition phenomena (see for example \cite{LL_92,MM1987,Wigg1988,Simo1999,BS_12}). The latter phenomena are associated with the dynamical nature of the motion fostered in the respective phase space, namely the regular or chaotic evolution of the system's orbits \cite{LL_92}.

The rapid and efficient detection of the regular or chaotic nature of motion in many dof systems has been a very active research topic over the years, starting with the theoretical introduction of Lyapunov Exponents (LEs) \cite{Lyapunov_1892,O_68}, and the development of efficient algorithms for their calculation \cite{BGGS_80a,BGGS_80b} (see also \cite{S10} for a recent review), which in time expanded and evolved as a subfield of dynamical systems leading to a gallery of chaos  detection methods (see e.g.~\cite{MDCG2011,DMCG2012,CMD2014} for comparisons among different methods and \cite{GotSkoLas2016} for a collection of recent reviews of several chaos indicators).

In this study, we employ the Generalized Alignment Index (GALI) \cite{SBA07}, a chaos detection method which has been shown to be a very efficient and fast tool for such purposes, especially in high dimensional problems (see e.g.~\cite{SBA08,BMC2009}). GALI's main advantages are (i) the quick distinction between regular and chaotic motion, (ii) the early detection of weak chaotic behavior, (iii) the determination of the  dimensionality of the torus on which the quasi-periodic motion occurs, and (iv) the prediction of slow diffusion \cite{SBA08}.

The GALI requires the evolution of several deviation vectors from a given orbit. The number $k$ of the vectors used defines the order of the index i.e.~GALI$_k$. In the case of regular motion GALI$_k$, for $k$ not greater than the dimension $N$ of the tangent space of the torus on which the motion takes place, eventually oscillates around a positive value, while for chaotic orbits it tends exponentially fast to zero \cite{SBA07}. In the case of stable periodic orbits of Hamiltonian systems, GALI$_k$ tends to zero following particular power laws, while for unstable periodic orbits it becomes zero exponentially fast \cite{MSA12}. A detailed discussion of the theory of GALIs, along with their applications to various dynamical systems can be found in \cite{SM16}.

In this paper, we seek out to investigate and understand the GALIs' prediction efficiency and performance when the initial condition (IC) of an  orbit moves gradually from a regular motion region to a chaotic one in Hamiltonian systems, by studying in detail the variation of the asymptotic values of the indices. This transition can take place either by changing the total energy of the system or by moving the regular orbit's IC towards the edge of a stability island. In addition, we follow the time evolution of the deviation vectors needed for the computation of the GALIs and examine if different initial distributions of the coordinates of these vectors affect the final GALI value. We perform our investigations in phase space regions around some simple, stable periodic orbits of the Fermi-Pasta-Ulam-Tsingou (FPUT) model \cite{FPUT55,F92}, which describes a chain of harmonic oscillators coupled through nonlinear interactions. The dynamics of this system has been studied extensively in the last decades (see e.g.~\cite{CamRosZas2005,Gal2008} and references therein) and is considered nowadays a prototypical, multidimensional, nonlinear model.

The paper is organized as follows. In Sect.~\ref{sec:model} we present the FPUT $\beta$ system, discuss the numerical integration of its equations of motion and its variational equations, and briefly present the definition of the GALI method, its properties, as well as its computation. Then, in Sect.~\ref{sec:gali_behavior} we study in detail the asymptotic behavior of the GALIs for regular orbits in the phase space neighborhood of two  types of stable periodic orbits of the FPUT $\beta$ model considering systems with different dof. In Sect.~\ref{sec:devvec} we focus our attention on the properties of the deviation vectors used for the computation of the GALIs, while in Sect.~\ref{sec:sum} we summarize our results and discuss our findings.

\section{Model and numerical techniques}\label{sec:model}

In our study, we consider the one-dimensional (1D) FPUT $\beta$ model~\cite{FPUT55,F92}, which represents a lattice chain of $N$ identical
particles with quartic nearest-neighbor interactions. The system's Hamiltonian function is given by
\begin{equation}
H_N = \sum_{i=1}^{N}\frac{1}{2}p_i^2 + \sum_{i=0}^{N} \left[ \frac{1}{2} (x_{i+1}-x_i)^2 +\frac{\beta}{4} (x_{i+1}-x_i)^4 \right],
\label{eq:FPUT_B}
\end{equation}
where $x_i$ is the displacement of the $i$-th particle from its equilibrium position and $p_i$ is the corresponding conjugate
momentum. In our investigation we impose fixed boundary conditions to system \eqref{eq:FPUT_B}, i.e.~$x_0=x_{N+1}=p_0=p_{N+1}=0$ and set $\beta=1$. The time evolution of a phase space orbit with ICs at time $t=t_0$ $\vec{z}(t_0)=(\vec{x}(t_0),\vec{p}(t_0))$, where $\vec{x}(t_0)=( x_1(t_0), x_2(t_0), \ldots, x_N(t_0))$ and $\vec{p}(t_0)=( p_1(t_0), p_2(t_0), \ldots, p_N(t_0))$,  is governed by the Hamilton equations of motion $\dot{x_i} =\frac{d x_i}{dt}= \frac{\partial H_N}{\partial p_i}$, $\dot{p_i} =\frac{d p_i}{dt}= -\frac{\partial H_N}{\partial x_i}$, $i=1,2,\ldots, N$.

In order to compute the GALI$_k$ chaos indicator for a particular orbit we have to follow the evolution of the orbit itself, along with a set of $k$, initially linearly independent, deviation vectors $\vec{v}_i$, $i=1,2,\ldots, k$, corresponding to $k$ different perturbations of the studied orbit. The time evolution of an initial deviation vector
$\vec{v}(t_0)=\vec{\delta z}(t_0)=(\vec{\delta x}(t_0),\vec{\delta p}(t_0))=(\delta x_1(t_0),\ldots,\delta x_N(t_0), \delta p_1(t_0), \ldots, \delta p_N(t_0))$ from an  orbit with ICs $\vec{z}(t_0)=(\vec{x}(t_0),\vec{p}(t_0))$ is governed by
the so-called variational equations (see e.g.~\cite{S10})
\begin{equation}
\label{eq:FPUT_var}
\dot{\vec{v}}(t) = \left[\textbf{J}_{2N} \cdot \textbf{D}_{H}^{2}(z(t))\right] \cdot \vec{v}(t_0),
\end{equation}
where $\textbf{J}_{2N}=
\bigl[ \begin{smallmatrix}
\textbf{0}_N & \textbf{I}_N \\
-\textbf{I}_N & \textbf{0}_N
\end{smallmatrix} \bigr] $,
with $\textbf{I}_N$ and $\textbf{0}_N$ being respectively the identity and the zero $N\times N$ matrices, and $\textbf{D}_{H}^{2}(z(t))$ is the
$2N\times 2N$ Hessian matrix with elements $\textbf{D}_{H}^{2}(z(t))_{i,j}=\frac{\partial^2 H}{\partial z_i \partial z_j}\Big|_{z(t)}, \quad i,j=1,2,\dots, 2N$.

Since Hamiltonian \eqref{eq:FPUT_B} can be split in two integrable parts $H_N(\vec{x},\vec{p})= A(\vec{p}) + B(\vec{x})$, with $A(\vec{p})$ being the kinetic energy which is  a function of only the momenta $p_i$, and $B(\vec{x})$ being the potential energy depending only on the coordinates $x_i$, we implement an efficient fourth-order symplectic integration scheme called
ABA864 \cite{BCFLMM13,SS18,DMMS19} for integrating the system's equations of motion. Symplectic integrators are numerical schemes specifically designed to preserve the symplectic structure of Hamiltonian systems. One of their main advantages is that they keep the error in the computed value of the Hamiltonian (which is typically referred to as the system's `energy') bounded for all times. In our numerical simulations, we typically integrate system \eqref{eq:FPUT_B} up to a final time $t_f=10^8$.  By adequately adjusting the used integration time step we always keep the absolute relative energy error below $10^{-8}$. By implementing the so-called `tangent map' technique \cite{SG10,GS11,GES12} we use the ABA864 scheme for integrating also the system's variational equations \eqref{eq:FPUT_var}, following in this way the time evolution of the set of deviation vectors needed for the computation of the GALIs.

\subsection{The GALI method}\label{sec:gali}

Let us now briefly discuss the GALI method and its properties considering our $N$ dof autonomous Hamiltonian \eqref{eq:FPUT_B}. By following an orbit $\vec{z}(t)=(\vec{x}(t),\vec{p}(t))$ and a set of $k$ initially independent deviation vectors $\vec{v}_i(t)$, $i=1,2,\ldots, k$, with $2\leq k \leq 2N$, the Generalized Alignment Index of order $k$ (GALI$_k$) represents  at any time $t$ the volume of the generalized parallelogram having as edges the $k$ unit deviation vectors $\hat{v}_i(t) = \frac{\vec{v}_i(t)}{\Vert \vec{v}_i(t) \Vert}$ \cite{SBA07}. This volume is computed
as the norm of the wedge product of these vectors
\begin{equation}
\mbox{GALI}_k(t) = \Vert \hat{v}_1(t) \wedge \hat{v}_2(t) \wedge  \ldots \wedge \hat{v}_k(t) \Vert.
\label{eq:GALI}
\end{equation}
We note that the number $k$ of the considered deviation vectors should
not exceed the dimension $2N$ of the system's phase space, because in that case the vectors will be by default linearly depended, and the corresponding volume (and consequently the value of GALI$_k$) will be zero. The GALI is a generalization of the Smaller Alignment Index (SALI) \cite{S01,SABV03,SABV04}, with GALI$_2$ being equivalent to SALI \cite{SBA07}. Both indices have been proven to be very efficient chaos indicators and have been successfully implemented in studies of various dynamical systems (see e.g.~\cite{SM16} and references therein).

In an $N$ dof Hamiltonian system regular motion typically occurs on $N$-dimensional ($N$D) tori. The system's dynamics lead any deviation vector of the regular orbit to eventually fall on the $N$D tangent space of the torus  \cite{SBA07}. Thus, asymptotically the volume defined by $k$ initially linearly independent deviation vectors, with $N < k \leq 2N$, will vanish as these vectors will become linearly dependent, while, in general, for $2 \leq k \leq N$ the volume will not become zero as the vectors will remain linearly independent. The general behavior of the GALI$_k$ for regular orbits lying on an $N$D torus of an $N$ dof Hamiltonian system is given by  \cite{SBA07}
\begin{equation}
\label{eq:GALI_reg}
\mbox{GALI}_k (t) \propto \left\{ \begin{array}{lll} \mbox{constant} &
 \mbox{if} &  2 \leq k \leq N  \\
  & & \\
 t^{-2(k-N)} & \mbox{if} & N< k \leq
    2N . \\
\end{array} \right.
\end{equation}
We note that the power law decay in \eqref{eq:GALI_reg} changes in the case of motion on a low-dimensional torus \cite{SBA08} (i.e.~a torus with dimensionality lower than $N$), and if some of the initial deviation vectors are already on the torus' tangent space \cite{SBA07}, but we will not discuss such cases here. Tori of regular motion exist around stable periodic orbits in conservative (or Hamiltonian) systems. For such orbits GALI$_k$ always decreases to zero following some specific power laws \cite{MSA12}
\begin{equation}
\label{eq:GALI_st}
\mbox{GALI}_k (t) \propto \left\{ \begin{array}{lll} t^{-(k-1)} &
 \mbox{if} &  2 \leq k \leq 2N-1  \\
  & & \\
 t^{-2N} & \mbox{if} &  k=2N. \\
\end{array} \right.
\end{equation}

On the other hand, for chaotic orbits and for unstable periodic orbits GALI$_k$ decays exponentially fast to zero \cite{SBA07,MSA12}
\begin{equation}
\label{eq:GALI_chaos}
\mbox{GALI}_k \propto e^{-[(\sigma_1-\sigma_2)+(\sigma_1-\sigma_3)+\dots+(\sigma_1-\sigma_k)]t},
\end{equation}
where $\sigma_1 \ge \sigma_2 \ge \dots \ge \sigma_k$ are approximations of the first $k$ largest LEs of the orbit.

An efficient way of computing the value of GALI$_k$ is through the Singular Value Decomposition (SVD) of the matrix
\begin{equation}
\label{eq:mat_dev}
\mathbf{A}=
\begin{pmatrix}
\hat{v}_{1} & \hat{v}_{2} & \cdots & \hat{v}_{k}\\
\end{pmatrix}
=
\begin{pmatrix}
\hat{v}_{1,1} & \hat{v}_{2,1} &  \cdots & \hat{v}_{k,1}\\
\hat{v}_{1,2} & \hat{v}_{2,2} &  \cdots & \hat{v}_{k,2}\\
\vdots & \vdots & \quad & \vdots \\
\hat{v}_{1,2N} & \hat{v}_{2,2N} &  \cdots & \hat{v}_{k,2N}\\
\end{pmatrix}
\end{equation}
having as columns the coordinates of the $k$ unitary vectors $\hat{v}_i(t) = \frac{\vec{v}_i(t)}{\Vert \vec{v}_i(t) \Vert}=\left( \hat{v}_{i,1}, \hat{v}_{i,2},\dots,\hat{v}_{i,2N} \right)$ \cite{SBA08}. In particular
\begin{equation}
\label{eq:SVD}
\mbox{GALI}_k=\prod_{i=1}^{k} z_i
\end{equation}
where $z_i$, $i=1,\dots,k$, are the so-called singular values of $\mathbf{A}$, obtained through
the SVD procedure.

We compute the value of GALI$_k$ by following the time evolution of $k$ initially independent random unit deviation vectors $\hat{v}_{1}(0), \hat{v}_{2}(0),\dots,\hat{v}_{k}(0)$. In order to statistically analyze the behavior of GALIs we average the indices over several different choices of the set of initial deviation vectors. The random choice of the initial vectors leads to different GALI$_k(0)$ values. Thus, in order to fairly and adequately compare the behavior of the indices for different initial sets of vectors we normalize the GALIs evolution by registering the ratio $\mbox{GALI}_k(t)/\mbox{GALI}_k(0)$, i.e.~we measure  the change of the volume defined by the $k$ deviation vectors with respect to the initially defined volume. Another option is to start the evolution of the dynamics by considering a set of $k$ \emph{orthonormal} vectors so that GALI$_k(0)=1$. As we will see later on, both approaches lead to similar results, so in our study we will follow the former procedure, unless otherwise stated.

\begin{figure}
\includegraphics[width=0.5\textwidth,keepaspectratio]{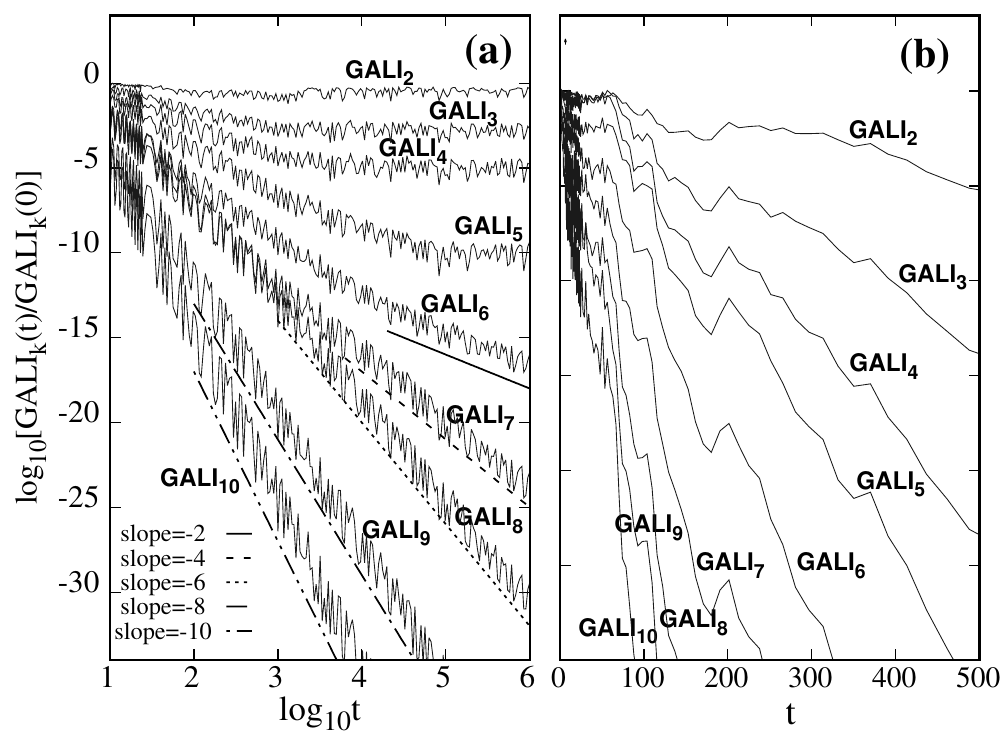}
  \caption{The time evolution of the normalized GALI indices $\mbox{GALI}_k(t)/\mbox{GALI}_k(0)$, $k=2,3,\ldots,10$ for (a) a regular orbit with ICs $x_1=-1.03003$, $x_3=-x_5=1.04003$, $p_1=0.29284$, and (b) a chaotic orbit with ICs $x_1=1.00097$, $x_3=-x_5=1.04003$, $p_1=0.57060$, of system \eqref{eq:FPUT_B} with $N=5$ and total energy $H_5=5$. All the other coordinates of both orbits are initially set to zero. The straight lines in (a) correspond to functions proportional to $t^{-2}$, $t^{-4}$, $t^{-6}$, $t^{-8}$ and $t^{-10}$, as indicated in the panel's legend. Both axes of (a) and the vertical axis of (b) are logarithmic. }
\label{fig:basic_GALI}
\end{figure}

In order to illustrate the basic behavior of the GALIs for regular and chaotic orbits we present in Fig.~\ref{fig:basic_GALI}(a) [Fig.~\ref{fig:basic_GALI}(b)] the time evolution of GALIs for a regular [chaotic] orbit of Hamiltonian \eqref{eq:FPUT_B} with $N=5$. The regular orbit of Fig.~\ref{fig:basic_GALI}(a) lies on a 5D torus and consequently GALI$_2$, GALI$_3$,
GALI$_4$ and GALI$_5$ eventually oscillate around some constant non-zero value, which decreases with increasing $k$. One of our main aims in this work is to analyze in depth this behavior. Note that the saturation of the indices to their practically constant limiting values happens at latter times for higher $k$ values. The GALIs with $k\geq 6$ eventually tend to zero following asymptotic power laws, which are indicated by straight lines in Fig.~\ref{fig:basic_GALI}(a). All these behaviors are in accordance to Eq.~\eqref{eq:GALI_reg}. On the other hand, as expected from \eqref{eq:GALI_chaos} the GALIs decay exponentially fast to zero in the case of the chaotic orbit of Fig.~\ref{fig:basic_GALI}(b). For the sake of completeness,  in Fig.~\ref{fig:basic_GALI}(b)
we also demonstrate the behavior of GALIs for
chaotic motion. Since we focus here on the behavior of GALIs for regular orbits, we do not investigate further the case of Fig.~\ref{fig:basic_GALI}(b), by e.g.~computing the corresponding LEs. Such investigations were systematically performed, for example, in \cite{SBA07,SBA08}.

\section{The behavior of the GALI for regular orbits}\label{sec:gali_behavior}

In general, regular motion occurs in the vicinity of stable periodic orbits in Hamiltonian systems. Thus, in order to study the behavior of the GALIs for regular orbits we first locate some stable periodic orbits of system \eqref{eq:FPUT_B} and then compute the GALIs for orbits in the neighborhood of these orbits.
The stability type of a periodic orbit is determined by
the eigenvalues of the so-called monodromy matrix $\mathbf{M}(T)$, which is obtained from the solution of the variational equations of the periodic orbit for one period $T$ (for more details, explicit equations and explanations see e.g.~\cite{S01b} and references therein).
The monodromy matrix is symplectic, and its columns are linearly independent solutions of the equations that govern the evolution of deviation vectors from the periodic orbit. If all the eigenvalues of $\mathbf{M}(T)$ are on the unit circle in the complex plane
the corresponding periodic orbit is stable, while otherwise it is unstable. We note that there exist several different types of instabilities \cite{S01}, but we will not discuss this issue here.

In an $N$D autonomous Hamiltonian system, two eigenvalues are always equal to $\lambda=1$ \cite{S01}, which means that in practice the remaining $2(N-1)$ eigenvalues define the periodic orbit's stability. Thus, we can reduce our investigation to a $2(N-1)$D subspace of the whole phase space through the well-known method of the Poincar{\'e} Surface of Section (PSS) (see e.g.~\cite{LL_92}), where the corresponding monodromy matrix has $2(N-1)$ eigenvalues, none of which is by default $\lambda=1$.

In what follows we investigate the behavior of the GALIs for regular orbits in the neighborhood of two simple periodic orbits (SPOs) of system \eqref{eq:FPUT_B}, which we refer to as  SPO1 and SPO2. The dynamics of these orbits was discussed in \cite{ABS06,AB06}.

\subsection{Regular motion in the neighborhood of SPO1}
\label{sec:SPO1}

The first SPO we study is called SPO1 in \cite{AB06} and it is obtained by considering an FPUT $\beta$ lattice \eqref{eq:FPUT_B} with $N$ being an odd integer, so that all particles at even-numbered positions are kept stationary at all times, while the odd-numbered particles are always displaced symmetrically to each other, i.e.
\begin{equation}
\label{eq:SPO1}
\hat{x}_{2j}(t)=0, \quad \hat{x}_{2j-1}(t)=-\hat{x}_{2j+1}(t)=\hat{x}(t),
\end{equation}
for $j=1,2, \ldots, \frac{N-1}{2}$. By inserting conditions  \eqref{eq:SPO1} in the system's equations of motion we end up  with a second order nonlinear differential equation for variable $\hat{x}(t)$ which describes the oscillations of all moving particles of SPO1. In particular,  for $j = 1,3,5,\ldots ,N$ we have
\begin{equation}
\label{eq:SPO1_difeq}
\ddot{\hat{x}}_j(t) = -2\hat{x}(t) - 2\beta\hat{x}^3(t),
\end{equation}
while $\hat{x}_j(t)=0$, $j= 2 , 4 , 6 , \ldots , (N-1)$ for all times.

The stability analysis of the SPO1  \cite{ABS06,AB06} showed that for small values of $H_N$ \eqref{eq:FPUT_B} the orbit is stable, but it becomes unstable when the energy increases beyond a certain threshold $H_N^c$. More specifically, in \cite{ABS06} it was shown that the energy density threshold   $H_N^c/N$ decreases with increasing $N$ following an asymptotic law $H_N^c/N \propto N^{-1}$ (see for example Fig.~1 of \cite{ABS06}).

\begin{figure}
\includegraphics[width=0.5\textwidth,keepaspectratio]{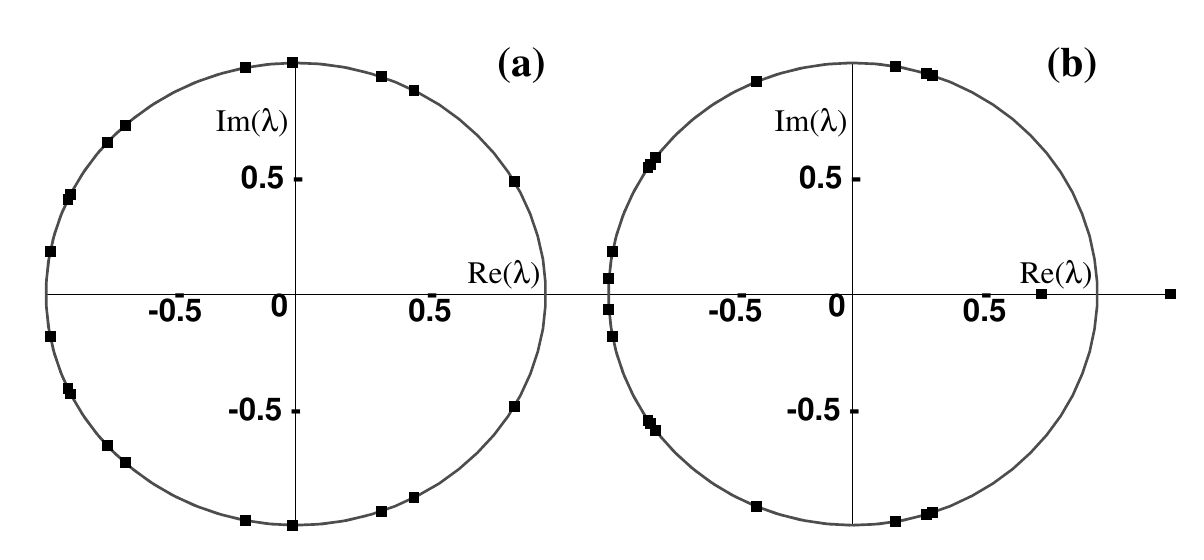}
  \caption{The arrangement on the complex plane of the eigenvalues $\lambda_i$, $i=1,2,\ldots,20$, of the monodromy matrix of the SPO1 \eqref{eq:SPO1} of Hamiltonian \eqref{eq:FPUT_B} with $N=11$ for (a) the stable SPO1 with $H_N/N=0.1$ and $\hat{x}(0)=-0.4112$, and (b) the unstable SPO1 with $H_N/N=0.2$ and $\hat{x}(0)=-0.5626$. The monodromy matrix is evaluated on the PSS (a) $x_1=-0.411$, $p_1>0$, and (b) $x_1=-0.560$, $p_1>0$. The critical energy density for which the SPO1 encounters its first transition  from stability to instability is $H_N^c/N \approx 0.176$.}
\label{fig:SPO1_eigen}
\end{figure}

In Fig.~\ref{fig:SPO1_eigen} we see the arrangement of the  eigenvalues  of the monodromy matrix of the SPO1 orbit  of Hamiltonian \eqref{eq:FPUT_B} with $N=11$ when $H_N/N=0.1$ [Fig.~\ref{fig:SPO1_eigen}(a)] and $H_N/N=0.2$ [Fig.~\ref{fig:SPO1_eigen}(b)]. In the first case the SPO1 is stable as all eigenvalues are on the unit circle, while in the latter the orbit is unstable as two eigenvalues are off the unit circle. We note that the transition from stability to instability for the SPO1 happens at    $H_N^c/N \approx 0.1755$.

To investigate how GALIs of different orders $k$ behave for regular orbits we perturb the stable SPO1 with ICs $\hat{x}_i(0)$, $\hat{p}_i(0)$, $i=1,2,\ldots, N$ and energy density $H_N/N$ to obtain a nearby orbit with ICs $X_i(0)$, $P_i(0)$, ensuring that the new orbit is at the same energy density level. The phase space distance $D$ between these two orbits in this case is given by
\begin{equation}
\label{eq:D}
D = \left\{\sum_{i=1}^{N} \left[ (\hat{x}_i(0) - X_i(0))^2 + (\hat{p}_i(0) - P_i(0))^2\right]
	 \right\}^{1/2}.
\end{equation}
Having this regular orbit lying on a $N$D torus means that GALI$_k$ with $2 \leq k \leq N$ will asymptomatically approach some almost constant value [see Eq.~\eqref{eq:GALI_reg}]. In order to statistically analyze these asymptotic  GALI$_k$ values we follow the evolution of GALIs for $n_v=10$ different random sets of initial deviation vectors whose coordinates are drawn from a uniform distribution in the interval $[-0.5, 0.5]$, and compute the evolution of the average (over the $n_v$ sets of deviation vectors) values $\langle \mbox{GALI}_k(t) / \mbox{GALI}_k(0) \rangle$. The integration of the orbit and of the deviation vectors is performed until the GALI values show a clear saturation. Then we estimate the asymptotic GALI value by finding the mean value of  $\langle \mbox{GALI}_k(t) / \mbox{GALI}_k(0) \rangle$ over the last $n_t=20$ recorded values, captured during, approximately, the one-fifth of the last decade of integration. We denote this quantity  as $\overline{ \mbox{GALI}}_k$ and estimate its error through the standard deviation of the considered  $n_t$  values.

\begin{figure}
\includegraphics[width=0.5\textwidth,keepaspectratio]{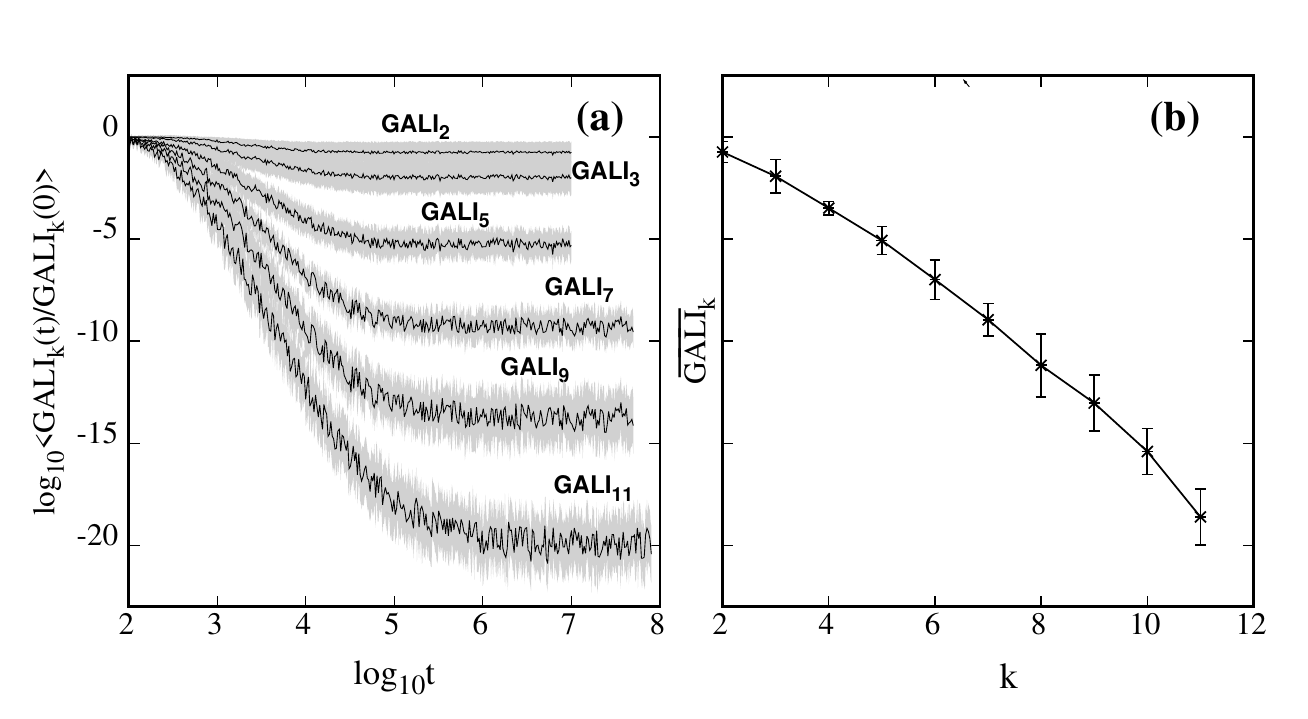}
  \caption{(a) The time evolution of the average, over $n_v=10$ sets of random initial deviation vectors, $\langle \mbox{GALI}_k(t) / \mbox{GALI}_k(0) \rangle$ (black curves) for $k=$2, 3, 5, 7, 9 and 11, for a regular orbit in the vicinity of the stable SPO1  with $H_N/N=0.01$ and $D=0.12$. The shaded gray area around each curve denotes 1 standard deviation. (b) Estimation of the asymptotic GALI values, $\overline{ \mbox{GALI}}_k$, as a function of their order $k$ for the orbit of panel (a). The error bars denote 1 standard deviation.  }
\label{fig:SPO1_11_GALIs}
\end{figure}

The outcome of this analysis for a regular orbit with distance \eqref{eq:D} $D=0.12$ from the stable SPO1 and $H_N/N=0.01$ is seen in Fig.~\ref{fig:SPO1_11_GALIs}. In particular, Fig.~\ref{fig:SPO1_11_GALIs}(a) depicts the evolution of $\langle \mbox{GALI}_k(t) / \mbox{GALI}_k(0) \rangle$ for some selected orders $k$ (black curves).The gray area around these curves represents one standard deviation. We observe that GALIs of higher orders converge to lower values and need more time to settle to these values. In  Fig.~\ref{fig:SPO1_11_GALIs}(b) we see more clearly how these final asymptotic values, $\overline{ \mbox{GALI}}_k$, decrease with increasing $k$.

\begin{figure}
\includegraphics[width=0.5\textwidth,keepaspectratio]{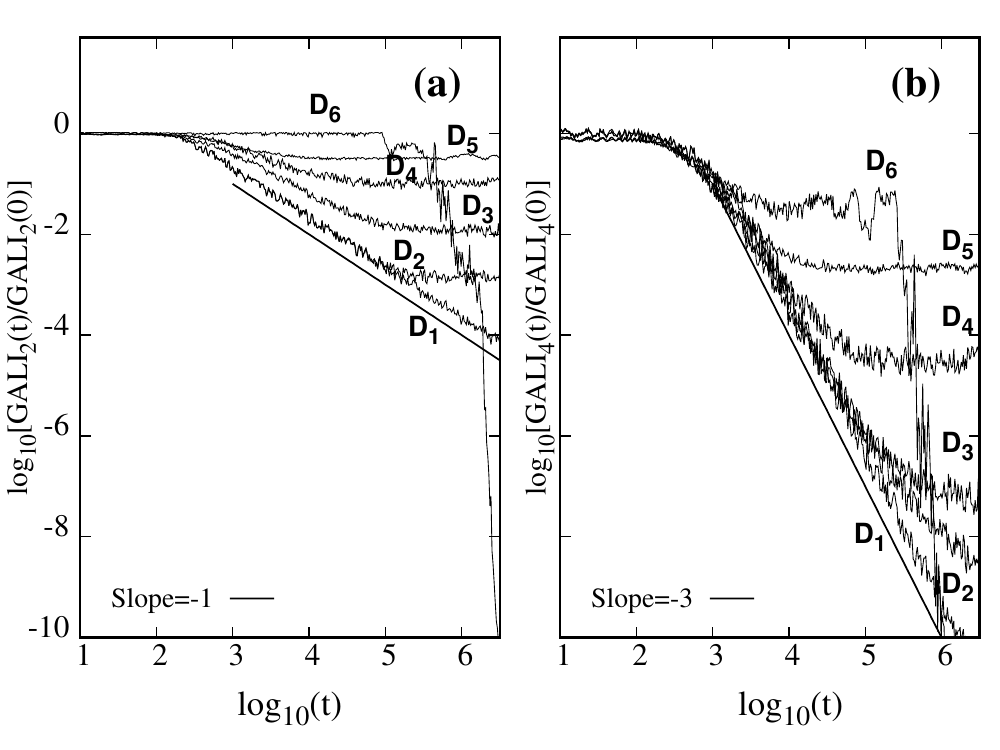}
  \caption{The time evolution of (a) $\mbox{GALI}_2(t) / \mbox{GALI}_2(0)$ and (b) $\mbox{GALI}_4(t) / \mbox{GALI}_4(0)$ for orbits with $H_N/N=0.01$ and distances $D_1 = 0.008$, $D_2 = 0.01$, $D_3 = 0.06$, $D_4 = 0.1$, $D_5 = 0.22$ and  $D_6 = 0.4$ from the stable SPO1,  for  1 set of initial deviation vectors. The straight lines  correspond to functions proportional to (a) $t^{-1}$, and (b) $t^{-3}$.  }
\label{fig:SPO1_11_change_D}
\end{figure}

Let us now investigate how these behaviors change when the studied regular orbit is taken further and further away from the stable SPO1, i.e.~as we increase $D$. A representative case is presented in Fig.~\ref{fig:SPO1_11_change_D} where we plot the evolution of GALI$_2$ [Fig.~\ref{fig:SPO1_11_change_D}(a)] and GALI$_4$   [Fig.~\ref{fig:SPO1_11_change_D}(b)] for several orbits in the neighborhood of the stable SPO1  with $H_N/N=0.01$ and $D_1 = 0.008$, $D_2 = 0.01$, $D_3 = 0.06$, $D_4 = 0.1$, $D_5 = 0.22$ and  $D_6 = 0.4$, when one set of random initial deviation vectors is used. First of all, we observe that the behavior of the GALIs for the regular orbit closest to the SPO1 ($D_1=0.008$) follows
the same power laws as the periodic orbit itself, namely \eqref{eq:GALI_st}, i.e.~GALI$_2 \propto t^{-1}$ and GALI$_2 \propto t^{-3}$. As $D$ increases, i.e.~the IC of the regular orbit moves further away from  the stable SPO1  inside the stability island surrounding the periodic orbit, GALIs start deviating from the power law decay observed for  $D=D_1$, and finally saturate to a positive value. This deviation starts earlier for larger $D$ values and consequently the asymptotic values of GALIs increase as $D$ grows. Eventually, for very large values of $D$, in the particular case discussed in Fig.~\ref{fig:SPO1_11_change_D} for $D=D_6=0.4$, the perturbed orbit becomes chaotic as it is located outside the stability island. In that case, the GALIs  decrease exponentially fast to zero in accordance with \eqref{eq:GALI_chaos}.

\begin{figure}
\includegraphics[width=0.4\textwidth,keepaspectratio]{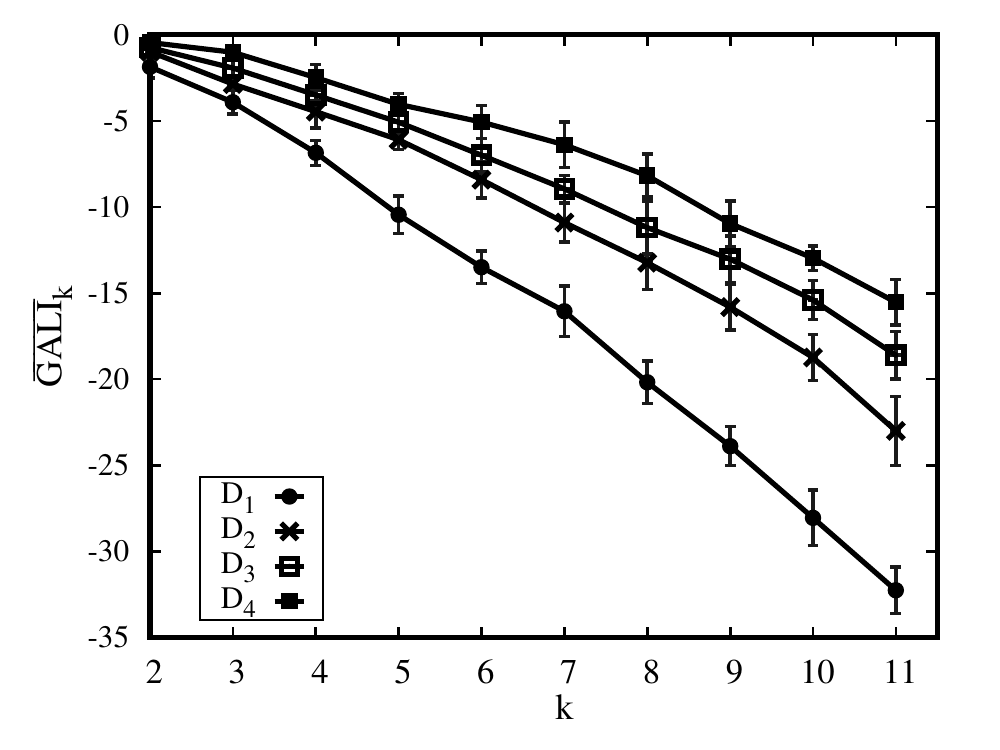}
  \caption{Similar to Fig.~\ref{fig:SPO1_11_GALIs}(b)  but for regular orbits in the vicinity of the stable SPO1 with $H_N/N=0.01$ and (a) $D_1=0.01$, $D_2=0.06$, $D_3=0.12$, and $D_4=0.22$.}
\label{fig:SPO1_11_change_D_many}
\end{figure}

In Fig.~\ref{fig:SPO1_11_change_D} we show the increase of only the asymptotic values of GALI$_2$ and GALI$_4$ with respect to $D$. This behavior is clearly seen for the whole spectrum of GALIs in Fig.~\ref{fig:SPO1_11_change_D_many} where we plot the  $\overline{ \mbox{GALI}}_k$ values versus $k$ for regular orbits around the SPO1 orbit with $H_N/N=0.01$ having $D$ values which are relatively not too small, in order to avoid the power law decays observed in Fig.~\ref{fig:SPO1_11_change_D} for $D=0.008$, but also relatively not too large in order for the studied orbit to be well inside the stability island.

\begin{figure}
\includegraphics[width=0.4\textwidth,keepaspectratio]{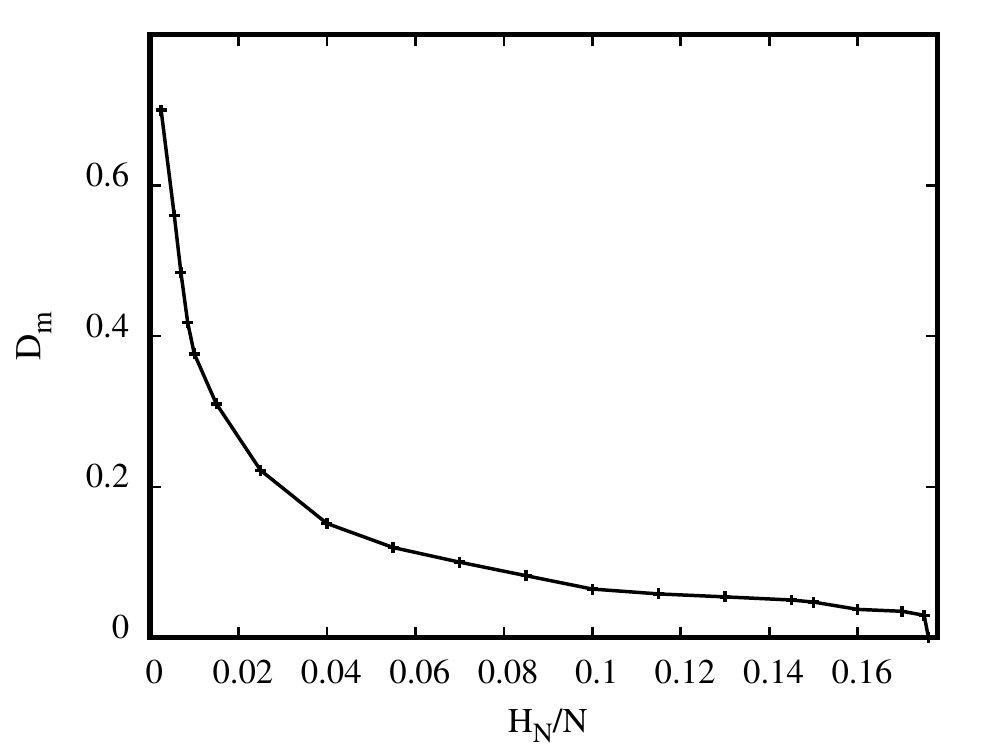}
  \caption{Estimation of the extent in phase space of the stability island around SPO1: The maximum value of $D$, denoted by $D_m$, for which regular motion occurs, as a function of the energy density $H_N/N$ for the Hamiltonian \eqref{eq:FPUT_B} with $N=11$.}
\label{fig:SPO1_D_energy}
\end{figure}

So far we saw how the asymptotic GALI values change when we increase the distance $D$ of the studied orbit from the stable SPO1 for constant energy $H_N$ (and obviously constant energy density $H_N/N$). In order to see how  $\overline{ \mbox{GALI}}_k$ values change when we increase the orbit's  energy density, we begin by estimating  the size of the stability island around the SPO1 orbit with different $H_N/N$ values by finding the largest $D$ value for which regular motion is observed. We denote this value $D_m$. The outcome of this process is shown in Fig.~\ref{fig:SPO1_D_energy}. The $D_m$ value (and consequently the size of the stability island) decreases as $H_N/N$ grows, and vanishes when the SPO1 periodic orbit destabilizes and becomes unstable. Recall that in this case the destabilization of the SPO1 takes place for  $H_N^c/N \approx 0.1755$. Thus, the region below the curve of Fig.~\ref{fig:SPO1_D_energy} corresponds to regular motion.

\begin{figure}
\includegraphics[width=0.4\textwidth,keepaspectratio]{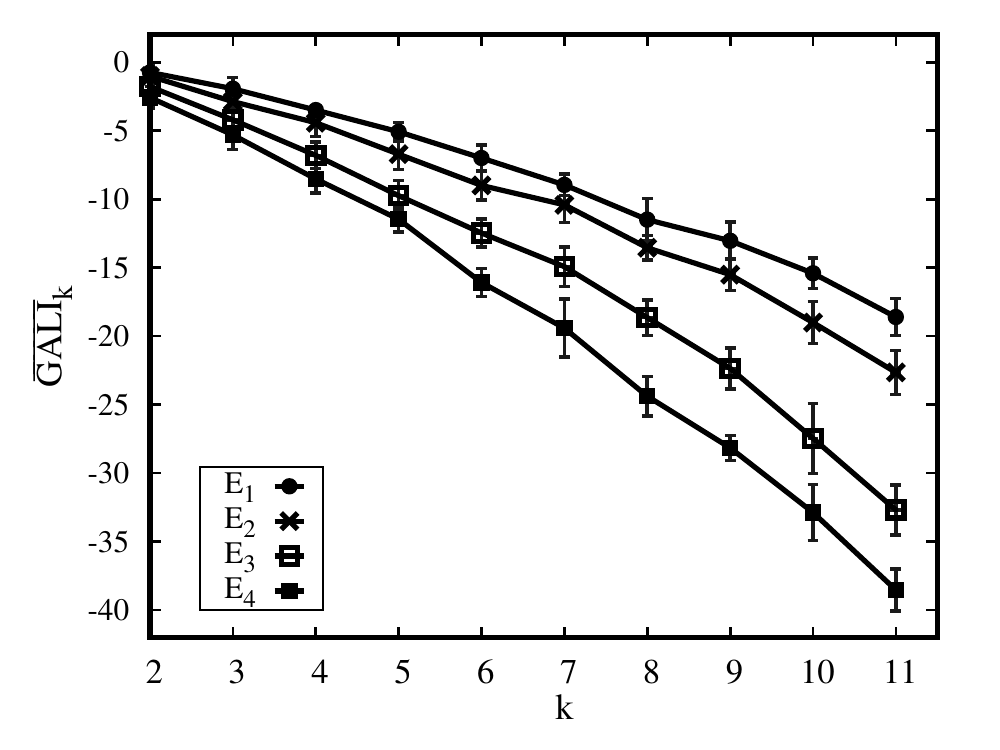}
  \caption{Similar to Fig.~\ref{fig:SPO1_11_change_D_many}  but for regular orbits with different energy densities  $E=H_N/N$: $E_1 = 0.01$, $E_2 = 0.04$, $E_3 = 0.12$ and  $E_4 = 0.16$. The phase space distance $D$ \eqref{eq:D} is  $D=0.1$ for $E_1$ and $E_2$, and $D=0.01$ for $E_3$ and $E_4$.}
\label{fig:SPO1_11_change_E_many}
\end{figure}

In Fig.~\ref{fig:SPO1_11_change_E_many} we present the values of $\overline{ \mbox{GALI}}_k$ as a function of the GALI's order $k$ for regular orbits in the neighborhood of the stable SPO1 for increasing energy densities $E=H_N/N$, i.e.~moving the orbit's IC in the region below the curve of Fig.~\ref{fig:SPO1_D_energy} towards the destabilization energy of SPO1. In particular, we consider regular orbits with $D=0.1$ for $E_1 = 0.01$, $E_2 = 0.04$ and with $D=0.01$ for $E_3 = 0.12$ and  $E_4 = 0.16$. As the energy density increases the regular region in Fig.~\ref{fig:SPO1_D_energy} is shrinking, thus the considered regular orbit is located closer and closer to the stable SPO1 orbit itself and the $D$ value should also decrease in order to avoid transition to chaotic motion. Since the regular orbit approaches the stable SPO1 the $\overline{ \mbox{GALI}}_k$ values shown in Fig.~\ref{fig:SPO1_11_change_E_many} decrease for orbits close to the periodic one, in accordance to what we observed in Fig.~\ref{fig:SPO1_11_change_D_many}.

\begin{figure}
\includegraphics[width=0.5\textwidth,keepaspectratio]{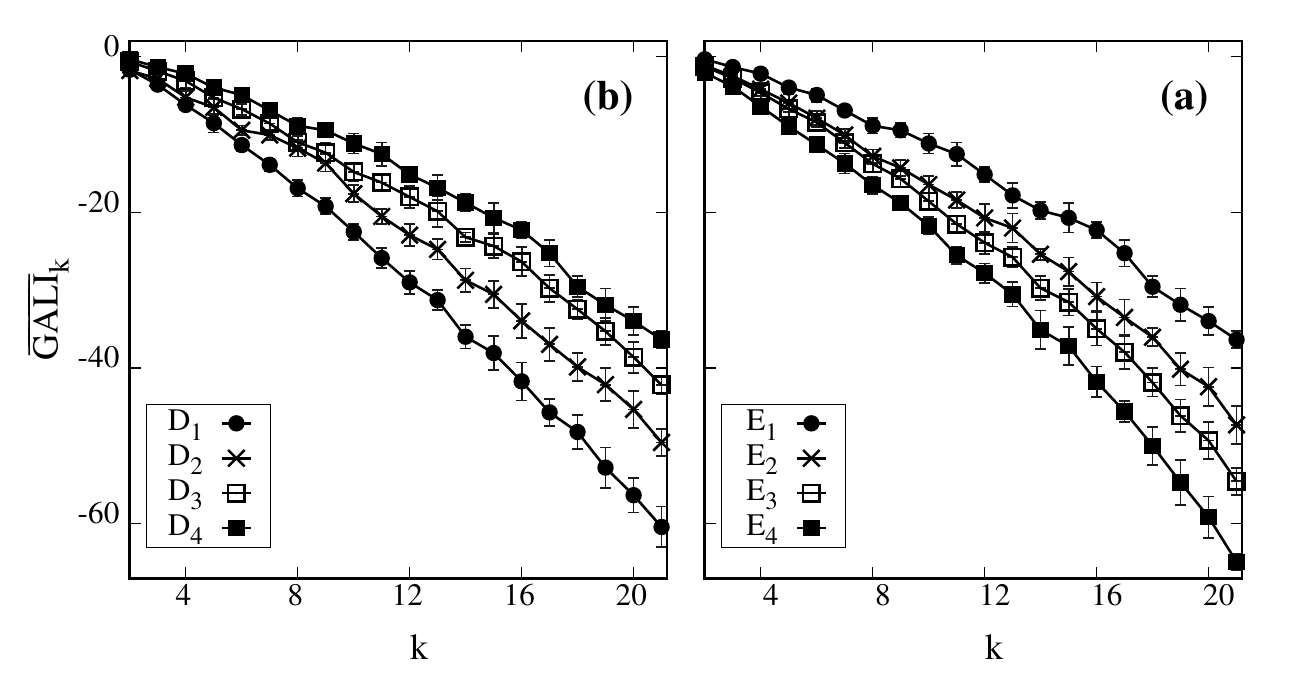}
  \caption{Estimation of the asymptotic  GALI values, $\overline{ \mbox{GALI}}_k$, as a function of their order $k$ for regular orbits in the vicinity of the stable SPO1 of system \eqref{eq:FPUT_B} with $N=21$. (a) Regular orbits for $E=H_N/N=0.002$ and distances $D$ \eqref{eq:D} $D_1=0.01$, $D_2=0.04$, $D_3=0.08$ and $D_4=0.12$. (b) Regular orbits with $E_1=0.002$ and $D=0.12$, $E_2=0.006$ and $D=0.1$, $E_3=0.01$ and $D=0.09$, and $E_4=0.04$ and $D=0.03$. The error bars in both panels denote 1 standard deviation. }
\label{fig:SPO1_21_change}
\end{figure}

The behavior we observe in the $N=11$ case for the asymptotic GALI values of regular orbits located further and further away from the stable SPO1 (Fig.~\ref{fig:SPO1_11_change_D_many}), as well as the one we see for regular orbits whose energies approach the destabilization energy of the SPO1 (Fig.~\ref{fig:SPO1_11_change_E_many}), remain the same also for other values of $N$. As a testimony to that we present in Fig.~\ref{fig:SPO1_21_change} results for the dependence of the   $\overline{ \mbox{GALI}}_k$ values on the index's order $k$ for the SPO1 orbit of system \eqref{eq:FPUT_B} with $N=21$. Note that in this case the destabilization energy density of the SPO1 is $H_N/N \approx 0.0675$.

\subsection{Regular motion in the neighborhood of SPO2}
\label{sec:SPO2}

The findings of Sect.~\ref{sec:SPO1} are not restricted to
only one type of SPO but are rather general. To demonstrate this we perform a similar analysis  for regular orbits in the neighborhood of what was called SPO2 in \cite{AB06}. This periodic orbit appears in FPUT $\beta$ systems \eqref{eq:FPUT_B} with $N=5+3 m$, $m=0,1,2,\ldots$, particles where every third particle remains always stationary and the two particles in between move in opposite directions
\begin{eqnarray}
    \label{eq:SPO2a}
	\hat{x}_{3j}(t)& = & 0, \quad j=1,2,3,\ldots ,\frac{N-2}{3}, \\
    \label{eq:SPO2b}
    \hat{x}_{j}(t)& = & -\hat{x}_{j+1}(t)=\hat{x}(t), \quad j=1,4,7,\ldots,N-1.
\end{eqnarray}
Similarly to the case of SPO1  inserting conditions \eqref{eq:SPO2a} and \eqref{eq:SPO2b} in the equations of motion of system \eqref{eq:FPUT_B} leads to a single differential equation  \begin{equation}
\label{eq:SPO1_difeq}
\ddot{\hat{x}}_j(t) = -3\hat{x}(t) - 9\beta\hat{x}^3(t),
\end{equation}
for the moving particles, while  $\hat{x}_j(t)=0$ for the stationary particles with $j=3,6,9,\ldots, N-2$. In \cite{AB06}, where the stability of this orbit was studied in detail, it was found that the destabilization energy density threshold $H_N^c/N$ decreases when the number of particles grows, as $H_N^c/N \propto N^{-2}$ (see Fig.~2(a) of \cite{AB06}).

\begin{figure}
\includegraphics[width=0.4\textwidth,keepaspectratio]{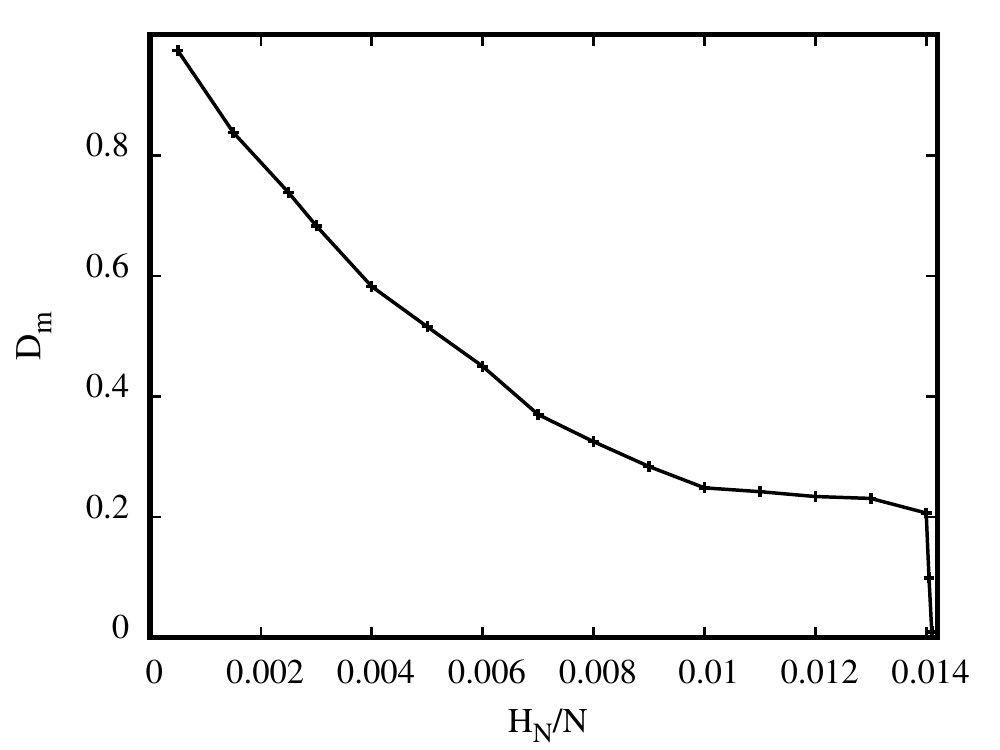}
  \caption{Similar to Fig.~\ref{fig:SPO1_D_energy} but for the SPO2 of Hamiltonian \eqref{eq:FPUT_B} with $N=11$.}
\label{fig:SPO2_D_energy}
\end{figure}

For $N=11$ the stability region of SPO2 is shown in Fig.~\ref{fig:SPO2_D_energy}, which can be directly compared with the stability region of SPO1 in Fig.~\ref{fig:SPO1_D_energy}, as both cases have the same number of particles. The shapes of both stability regions look similar but the extend of the SPO2 region in the energy density axis is much smaller as the stable SPO2 becomes unstable for   $H_N^c/N \approx 0.01395$.

\begin{figure}
\includegraphics[width=0.5\textwidth,keepaspectratio]{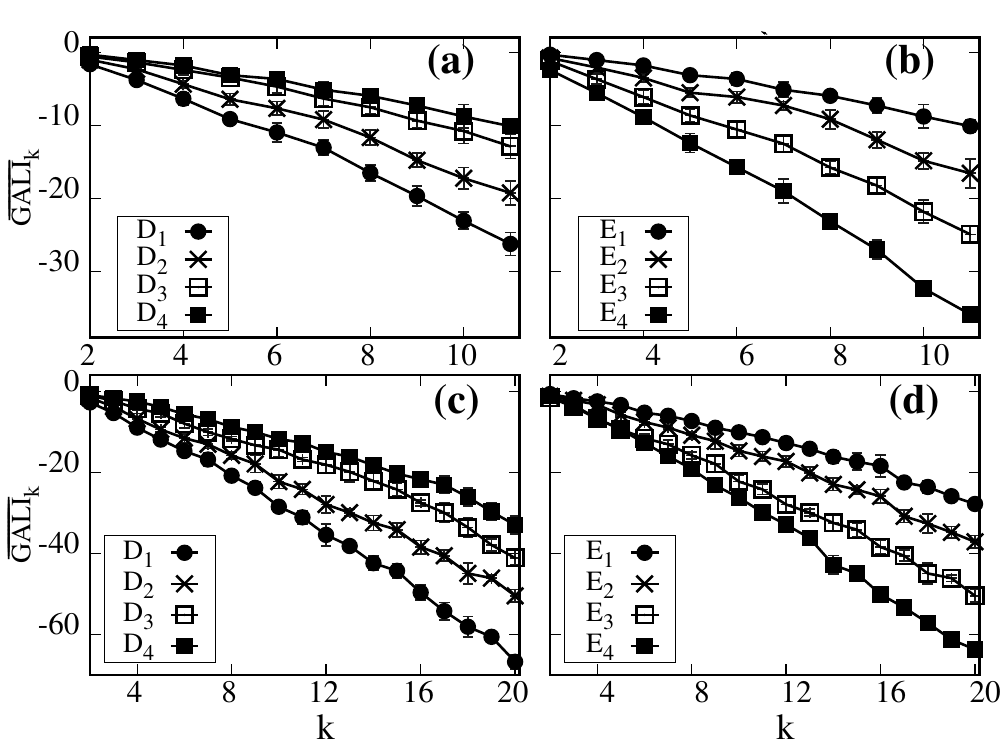}
  \caption{Estimation of the asymptotic GALI values, $\overline{ \mbox{GALI}}_k$, as a function of their order $k$ for regular orbits in the vicinity of the stable SPO2 of system \eqref{eq:FPUT_B} with $N=11$ [(a) and (b)] and $N=20$ [(c) and (d)]. The considered regular orbits have (a) $E=H_N/N=0.001$ and distances $D$ \eqref{eq:D} from the stable SPO2 $D_1=0.01$, $D_2=0.04$, $D_3=0.08$ and $D_4=0.12$, (c) $E=H_N/N=0.002$ with  $D_1=0.01$, $D_2=0.04$, $D_3=0.08$ and $D_4=0.12$, (b) $E_1=0.001$ and $D=0.12$, $E_2=0.004$ and $D=0.12$, $E_3=0.008$ and $D=0.04$, and $E_4=0.012$ and $D=0.01$, (d) $E_1=0.0004$ and $D=0.1$, $E_2=0.0008$ and $D=0.04$, $E_3=0.002$ and $D=0.04$, and $E_4=0.004$ and $D=0.002$.}
\label{fig:SPO2_11_20_change}
\end{figure}

In Fig.~\ref{fig:SPO2_11_20_change} we present results regarding  the asymptotic GALI values, $\overline{ \mbox{GALI}}_k$, for regular orbits inside the stability island around the SPO2 orbit for $N=11$ [Figs.~\ref{fig:SPO2_11_20_change}(a) and (b)] and for $N=20$ [Figs.~\ref{fig:SPO2_11_20_change}(c) and (d)]. We observe similar behaviors to the ones encountered in the neighborhood of SPO1. In particular, the $\overline{ \mbox{GALI}}_k$ values increase as we move away from the stable periodic orbit [Figs.~\ref{fig:SPO2_11_20_change}(a) and (c)], similarly to what we observed in Figs.~\ref{fig:SPO1_11_change_D_many} and \ref{fig:SPO1_21_change}(a), while they decrease as the energy of the regular orbit approaches the destabilization energy of SPO2 [Figs.~\ref{fig:SPO2_11_20_change}(b) and (d)] as we saw in Figs.~\ref{fig:SPO1_11_change_E_many} and \ref{fig:SPO1_21_change}(b) for the SPO1 case. We note that for $N=20$ the destabilization threshold is in this case $H_N^c/N \approx 0.00425$.

\section{Statistical analysis of deviation vectors}\label{sec:devvec}

In the case of chaotic orbits it is known that all, initially different, deviation vectors will eventually become aligned to the direction defined by the maximum LE \cite{BGGS_80a,BGGS_80b,S10}. This behavior leads to the eventual vanishing of the GALIs through the exponential decay \eqref{eq:GALI_chaos} \cite{SBA07}. On the other hand, for regular orbits we know that deviation vectors eventually fall on the tangent space of the torus on which the motion occurs \cite{SBA07}, but it is not clear how this behavior affect the actual GALI values. For this reason we  investigate here in  detail the behavior of  deviation vectors for regular orbits.

\begin{figure}
\includegraphics[width=0.5\textwidth,keepaspectratio]{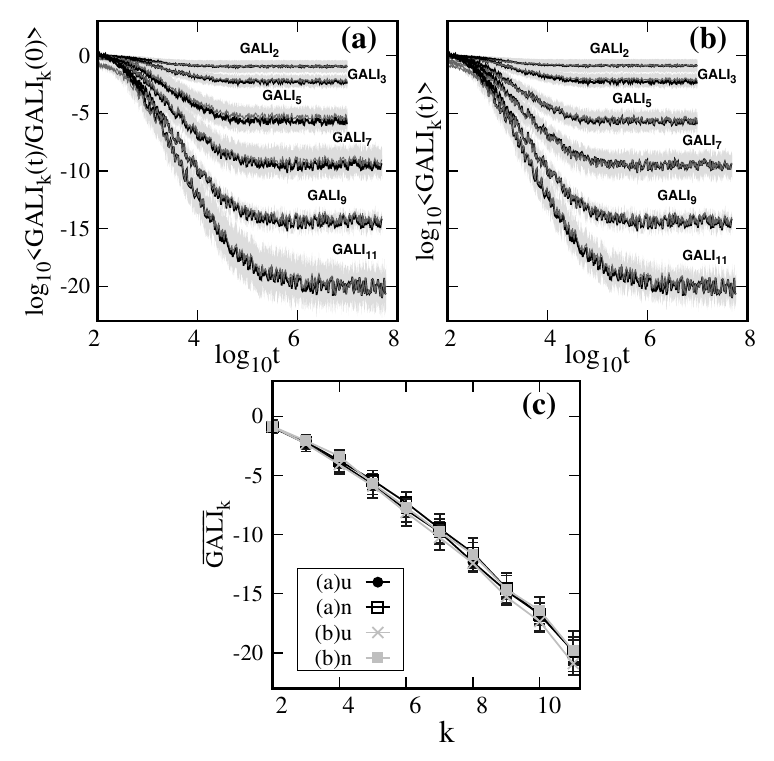}
  \caption{The time evolution of the average over $n_v=10$ sets of initial deviation vectors of (a) $\langle \mbox{GALI}_k(t) / \mbox{GALI}_k(0) \rangle$ and (b) $\langle \mbox{GALI}_k(t) \rangle$ for $k=2$, 3, 5, 7, 9, 11, for a regular orbit in the neighborhood of the stable SPO1 ($D=0.12$, $H_N/N=0.01$) of Hamiltonian \eqref{eq:FPUT_B} with $N=11$. The coordinates of the initially (a) unit  [(b) orthonormalized] deviation vectors were chosen from a uniform (black curves) and a normal distribution (light black curves). The 2 curves practically coincide in panels (a) and (b), while the gray area around them indicate 1 standard deviation. (c) Similar to Fig.~\ref{fig:SPO1_11_GALIs}(b) but for the 4 different cases of panels (a) and (b). The cases are denoted in the legend by the panels name, (a) or (b), and by `u' or `n' for uniform and normal distributions. The 4 curves practically overlap. }
\label{fig:dif_dev_vec}
\end{figure}

An important finding is that the  evolution of the GALIs does not practically depend on the initial choice of deviation vectors needed for their computation. A particular example illustrating this property is shown in Fig.~\ref{fig:dif_dev_vec} where we present the evolution of GALIs for a regular orbit close to the stable SPO1 ($D=0.12$) of Hamiltonian \eqref{eq:FPUT_B} for $N=11$ and $H_N/N=0.01$. In Fig.~\ref{fig:dif_dev_vec}(a) we show the evolution of $\langle \mbox{GALI}_k(t) / \mbox{GALI}_k(0) \rangle$ for some selected values of $k$ when the coordinates of the unit initial deviation vectors are chosen from a uniform distribution in the interval $[-0.5, 0.5]$ (black curves) or from a normal distribution with mean $0$ and standard deviation $1$ (light black curves). The two curves practically coincide. The presented results are averaged over $n_v=10$ random sets of initial deviation vectors and the gray area around the curves indicate one standard deviation, as was done for example in Fig.~\ref{fig:SPO1_11_GALIs}(a). In Fig.~\ref{fig:dif_dev_vec}(b) we show a similar computation, but now the initial deviation vectors, whose initial coordinates were created in a similar way as in Fig.~\ref{fig:dif_dev_vec}(a), were orthonormalized at the beginning of their evolution. This choice sets GALI$_k(0)=1$ and for this reason we report in Fig.~\ref{fig:dif_dev_vec}(b) the average value $\langle \mbox{GALI}_k(t) \rangle$. Also for this case we see that the method of creation of the initial, orthonormal deviation vectors does not affect the evolution of the GALIs. Furthermore, by comparing Figs.~\ref{fig:dif_dev_vec}(a) and (b) we observe that also the choice of unit or orthonormal initial deviation vectors does not have any practical influence on the evolution of the indices. This becomes more evident from the results of Fig.~\ref{fig:dif_dev_vec}(c) where we plot the estimations of the asymptotic GALI values, $\overline{ \mbox{GALI}}_k$, obtained from the results of Figs.~\ref{fig:dif_dev_vec}(a) and (b) [in the same way that Fig.~\ref{fig:SPO1_11_GALIs}(b) was obtained from Fig.~\ref{fig:SPO1_11_GALIs}(a)], as the four curves practically overlap.

\begin{figure}
\includegraphics[width=0.5\textwidth,keepaspectratio]{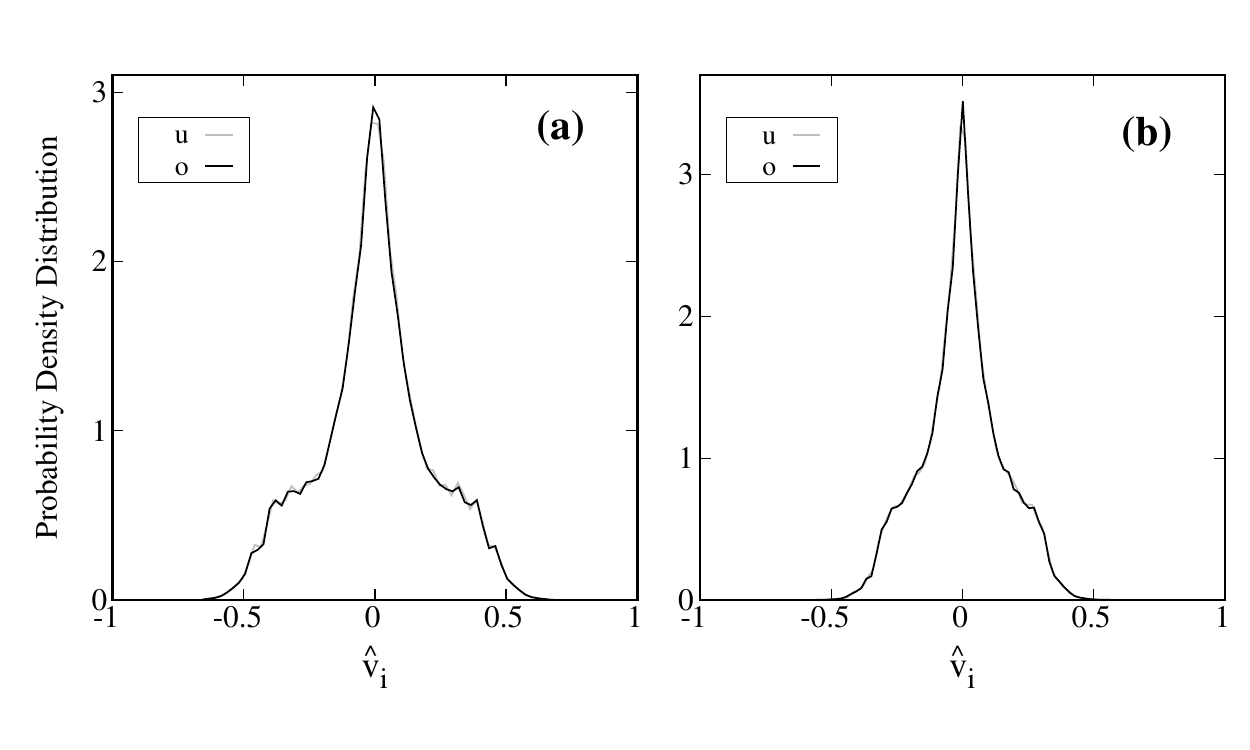}
  \caption{Probability density distributions of the coordinates $\hat{v}_i$ of 11 initially unit (black curves - `u') or orthonormalized (gray curves - `o') deviation vectors for a regular orbit of Hamiltonian \eqref{eq:FPUT_B} with $N=11$ close to the stable SPO1, with (a) $H_N/N=0.01$, $D=0.12$, and (b) $H_N/N=0.002$, $D=0.04$. The distributions are created from data obtained for 10 snapshots of the deviation vectors' evolution when the GALI$_{11}$ values have practically reached their asymptotic behavior. In both panels the two curves practically overlap.}
\label{fig:distributions_1}
\end{figure}

In Fig.~\ref{fig:distributions_1} we plot the probability density distributions of the coordinates of the unit deviation vectors needed for the evaluation of GALI$_{11}$ for a regular orbit close to SPO1 of  Hamiltonian \eqref{eq:FPUT_B} with $N=11$. In particular, we consider the orbit with $H_N/N=0.01$, $D=0.12$  in Fig.~\ref{fig:distributions_1}(a) and $H_N/N=0.002$, $D=0.04$ in Fig.~\ref{fig:distributions_1}(b). The distributions are created by the coordinates of the set of 11 deviation vectors obtained at 10 snapshots when the GALI$_{11}$ has reached  its asymptotic value. Black curves correspond to initially unit deviation vectors, while gray curves to orthonormal ones. Both curves practically overlap.

The results of Figs.~\ref{fig:dif_dev_vec}  and \ref{fig:distributions_1} clearly indicate that the initial distribution of the deviation vector coordinates does not play a role in the evolution of GALIs. Thus, we will continue our investigations by considering only initial unit deviation vectors having their coordinates generated from a uniform distribution.

\begin{figure}
\includegraphics[width=0.5\textwidth,keepaspectratio]{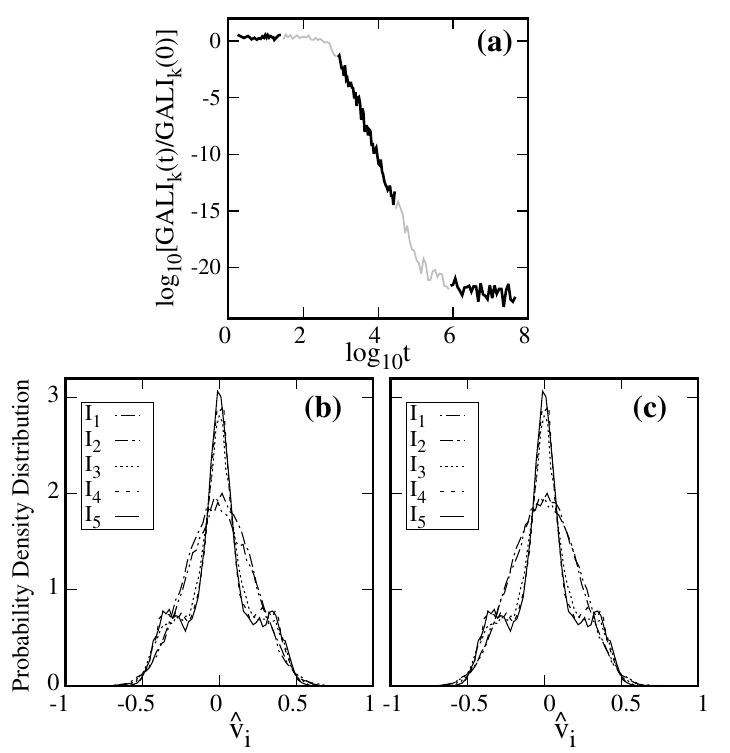}
  \caption{(a) The time evolution of GALI$_{11}$ for a regular orbit in the neighborhood of the stable SPO1 orbit of Hamiltonian \eqref{eq:FPUT_B} for $N=11$ with $H_N/N=0.004$ and $D=0.04$. The evolution is divided into 5 intervals, I$_1$: $0 \leq \log_{10}t < 1.5$,  I$_2$: $1.5\leq \log_{10}t < 3$, I$_3$: $3 \leq \log_{10}t < 4.5$, I$_4$: $4.5 \leq \log_{10}t < 6$ and I$_5$: $6 \leq \log_{10}t \leq 7.5$. The coordinate distributions of the 11 unit deviation vectors  for the intervals I$_1$, I$_2$, I$_3$, I$_4$ and I$_5$  are shown in (b) for the deviation vectors used in (a), while in (c) we see results similar to the ones of (b) but for another random set of initial deviation vectors.}
\label{fig:distribution_evolve}
\end{figure}

The distribution of the deviation vectors coordinates evolve in time as we see in Fig.~\ref{fig:distribution_evolve} where we divide into five time intervals the evolution of GALI$_{11}$ for a regular orbit close to the stable SPO1  of Hamiltonian \eqref{eq:FPUT_B} for $N=11$, with $H_N/N=0.004$ and $D=0.04$ [Fig.~\ref{fig:distribution_evolve}(a)]. Each interval has the same length in logarithmic scale. The corresponding coordinate distributions are shown in Fig.~\ref{fig:distribution_evolve}(b), while in Fig.~\ref{fig:distribution_evolve}(c) we present similar distributions but for another, random set of initial deviation vectors. From
Figs.~\ref{fig:distribution_evolve}(b) and (c) we  see the evolution of the distributions from a rather spread `triangular' shape to a more concentrated one, along with clear similarities between the distributions obtained by different initial sets of deviation vectors. Thus, in general, not only the final distribution obtained when GALIs have reached their asymptotic behavior, but also the time evolution of the coordinate distributions is practically independent of the initial choice of deviation vectors.

\begin{figure}
\includegraphics[width=0.5\textwidth,keepaspectratio]{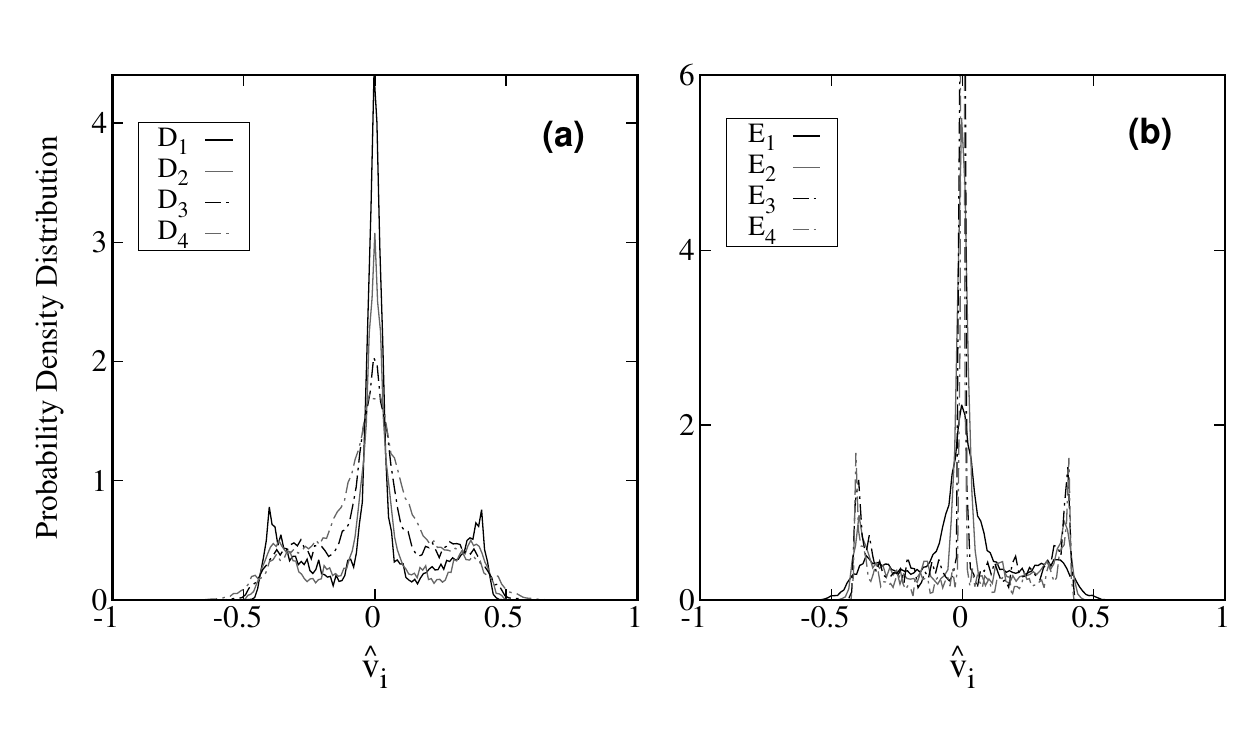}
  \caption{The final coordinate distributions of the 11 unit deviation vectors used for the computation of GALI$_{11}$ for regular orbits in the neighborhood of the stable SPO1 of Hamiltonian \eqref{eq:FPUT_B} with $N=11$. The considered regular orbits have (a) $E=H_N/N=0.01$ and distances $D$ \eqref{eq:D} from the stable SPO1 $D_1=0.01$, $D_2=0.06$, $D_3=0.12$ and $D_4=0.22$, and (b)  $E_1=0.001$ and $D= 0.015$, $E_2=0.01$ and $D= 0.015$, $E_3=0.1$ and $D= 0.015$, and $E_4=0.175$ and $D=0.001$.   }
\label{fig:distributions_ED_SPO1}
\end{figure}

\begin{figure}
\includegraphics[width=0.5\textwidth,keepaspectratio]{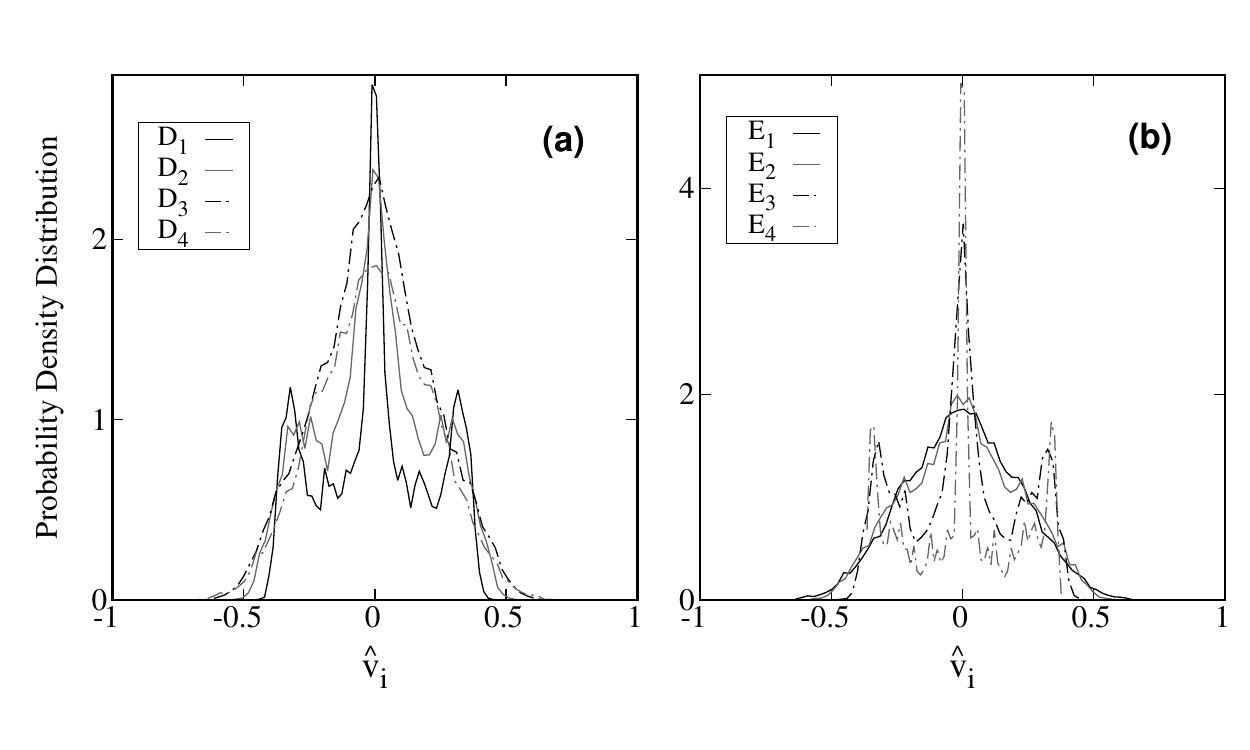}
  \caption{The final coordinate distributions of the 11 unit deviation vectors used for the computation of GALI$_{11}$ for regular orbits in the neighborhood of the stable SPO2 of Hamiltonian \eqref{eq:FPUT_B} with $N=11$. The considered regular orbits have (a) $E=H_N/N=0.001$ and distances $D$ \eqref{eq:D} from the stable SPO2 $D_1=0.01$, $D_2=0.04$, $D_3=0.08$ and $D_4=0.12$, and (b)  $E_1=0.001$ and $D= 0.12$, $E_2=0.004$ and $D= 0.12$, $E_3=0.008$ and $D= 0.04$, and $E_4=0.012$ and $D=0.01$.   }
\label{fig:distributions_ED_SPO2}
\end{figure}

In Figs.~\ref{fig:distributions_ED_SPO1} and \ref{fig:distributions_ED_SPO2} we investigate the dependence of the final deviation vector coordinate distribution on the regular orbit's distance $D$ from the stable SPO in the case of  SPO1 [Fig.~\ref{fig:distributions_ED_SPO1}(a)] and  SPO2 [Fig.~\ref{fig:distributions_ED_SPO2}(a)]. We see that in both cases we have a transition from a rather pointy distribution with high concentrations in the distribution's middle and edges (small $D$ values) to a more `triangular' shape as we approach the boundaries of the stability island (large $D$ values). In accordance to this behavior we see in Figs.~\ref{fig:distributions_ED_SPO1}(b) and \ref{fig:distributions_ED_SPO2}(b) that when the orbit's energy density increases towards the SPO's destabilization energy the distributions move to a more pointy and concentrated shape, because the regular orbit is located closer to the stable SPO, as we have already pointed out in the discussion of Fig.~\ref{fig:SPO1_11_change_E_many}.

\section{Summary and discussion}\label{sec:sum}

In this work we investigated in detail the behavior of the GALI chaos indicator for regular motion of multidimensional Hamiltonian systems. Thus, our study completes in some sense previous works on the GALI method \cite{SBA07,SBA08,MSA12} where other aspects of the index were investigated, like for example its behavior for periodic orbits \cite{MSA12}. In particular, we considered several regular orbits in the vicinity of two basic SPOs of the FPUT $\beta$ model \eqref{eq:FPUT_B}, the so-called SPO1 and SPO2 orbits \cite{AB06}, for various numbers of the system's dof.

We showed that the time evolution of the GALIs, along with the distribution of the coordinates of the deviation vectors needed for their computation, are not influenced by the initial choice of the  deviation vectors (Figs.~\ref{fig:dif_dev_vec} and  \ref{fig:distributions_1}). In order to do so we considered various sets of initial deviation vectors. More specifically, we used unit vectors whose coordinates were drawn randomly from a uniform or a normal distribution. In that case, since the initial value GALI$_k(0)$  is different for each choice of vectors, we registered the evolution of   $\langle \mbox{GALI}_k(t) / \mbox{GALI}_k(0) \rangle$. An alternative option we considered was vectors, whose coordinates were again randomly generated from a uniform or a normal distribution, which were also orthonormalized at the beginning of our numerical simulations, setting in this way  GALI$_k(0)=1$. For this choice we followed the evolution of $\langle \mbox{GALI}_k(t) \rangle$.

The value of GALI$_k$ eventually saturates to a positive value for regular motion on an $N$D torus for $2 \leq k \leq N$. The asymptotic $\overline{ \mbox{GALI}}_k$ values depend on (i) the order $k$ of the index, i.e.~GALIs of higher order attain lower values, (ii) the phase space distance $D$ \eqref{eq:D} of the regular orbit from the nearby SPO, i.e.~the $\overline{ \mbox{GALI}}_k$ values increase when $D$ grows, and (iii) the orbit's energy $H_N$ \eqref{eq:FPUT_B}, or equivalently the energy density $H_N/N$, i.e.~as we approach the destabilization energy of the SPO the   $\overline{ \mbox{GALI}}_k$ values decrease. Furthermore, we showed that also the shape of the coordinate distributions of the deviation vectors depend on $D$ and  $H_N/N$ (Figs.~\ref{fig:distributions_ED_SPO1} and \ref{fig:distributions_ED_SPO2}), although these distributions are not influenced by the initial choice of the vectors.

\acknowledgments
H.~M.~was partially funded by the University of Cape Town (UCT) International and Refugee Grant, as well as by the Woldia University and MoSHE. Ch.~S.~acknowledges support by the UCT's Research Committee (URC). We also thank the High Performance Computing facility of UCT (\url{http://hpc.uct.ac.za}) and the Center for High Performance Computing (\url{https://www.chpc.ac.za}) for  providing the computational resources needed for  this work.


\end{document}